\documentclass[preprint,aps]{revtex4}
\usepackage{amssymb,epsf}

\begin{document}

\title{The effects of nonlinear Maxwell source on the magnetic solutions in
Einstein-Gauss-Bonnet gravity}
\author{S. H. Hendi$^{1,2,3}$\thanks{%
email address: hendi@mail.yu.ac.ir}, S. Kordestani$^{1}$ and S. N.
Doosti Motlagh$^{1}$} \affiliation{$^1$ Physics Department,
College of Sciences, Yasouj University, Yasouj
75914, Iran\\
$^2$Research Institute for Astrophysics and Astronomy of Maragha
(RIAAM),
P.O. Box 55134-441, Maragha, Iran\\
$^3$National Elite Foundation, P.O. Box 19615-334, Tehran, Iran}

\begin{abstract}
Considering both the power Maxwell invariant source and the
Einstein--Gauss--Bonnet gravity, we present a new class of static
solutions yields a spacetime with a longitudinal nonlinear
magnetic field. These horizonless solutions have no curvature
singularity, but have a conic geometry with a deficit angle
$\delta \phi$. In order to have vanishing electromagnetic field at
spatial infinity, we restrict the nonlinearity parameter to
$s>1/2$. Investigation of the energy conditions show that these
solutions satisfy the null, weak and strong energy conditions
simultaneously, for $s>1/2$, and the dominant energy condition is
satisfied when $s \in \left( {\frac{1}{2},1}\right]$. In addition,
we look for about the effect of nonlinearity parameter on the
energy density and also deficit angle, and find that these
quantities are sensitive with respect to variation of nonlinearity
parameter. We find that for special values of nonlinearity
parameter, two important subclass of solutions, so-called
conformally invariant Maxwell and BTZ-like solutions, with
interesting properties, emerge. Then, we generalize the static
solutions to the case of spinning magnetic solutions and find
that, when one or more rotation parameters are nonzero, the brane
has a net electric charge which is proportional to the magnitude
of the rotation parameters. We also use the counterterm method to
compute the conserved quantities of these spacetimes such as mass,
angular momentum, and find that these conserved quantities do not
depend on the nonlinearity parameter.
\end{abstract}

\maketitle

\section{Introduction}
Among the theories of gravity with higher derivative corrections,
the Gauss-Bonnet (GB) gravity is quite special. Indeed, in order
to have a ghost-free action, the quadratic curvature corrections
to the Einstein-Hilbert action should not contain derivatives of
metrics of order higher than second, and should be proportional to
the GB term \cite{Boulware}. This combination also appear
naturally in the next-to-leading order term of the heterotic
string effective action, and plays a fundamental role in some
gravitational theories \cite{Cham}. Generally, in recent years, GB
gravity has been studied by many authors (see \cite
{Wil1,DH,Deh1,Deh2,DehMagGB,Deh3,Levi,Vil,Mukh,Dias,Dehghani,NUT,Od1}
and references therein).

In the conventional, straightforward generalization of the Maxwell
field to higher dimensions one essential property of the
electromagnetic field is lost, namely, conformal invariance. The
first black hole solution derived for which the matter source is
conformally invariant is the Reissner-Nordstr\"{o}m solution in
four dimensions. Indeed, in this case the source is given by the
Maxwell action which enjoys the conformal invariance in four
dimensions. Maxwell theory can be studied in a gauge which is
invariant under conformal rescalings of the metric, and firstly,
has been proposed by Eastwood and Singer \cite{EasSin}. Recently,
there exists a nonlinear extension of the Maxwell Lagrangian in
higher dimensions, if one uses the Lagrangian of the $U(1)$ gauge
field in the form
\cite{HasMar,HendiRastegar,HendiPLB,HendiEslamPanah}
\begin{equation}
L=F^{s},  \label{IGMF}
\end{equation}
where $F=F_{\mu \nu }F^{\mu \nu }$ is the Maxwell invariant,
$F_{\mu \nu }=\partial _{\mu }A_{\nu }-\partial _{\nu }A_{\mu }$
is the Maxwell tensor and $A_{\mu }$\ is the vector potential. In
what follows, we consider this Lagrangian as the matter source
coupled to the Einstein-GB gravity. The first motivation is to
take advantage of the conformal symmetry to construct the
analogues of the four-dimensional Reissner-Nordstr\"{o}m black
hole solutions in higher dimensions, and the second motivation
comes from the generalization of Maxwell field and investigation
of their effects on the energy-momentum tensor.

In this paper we want to restrict ourself at most to the first
three terms of Lovelock gravity. The first two terms are the
Einstein-Hilbert term with cosmological constant, while the third
term is known as the Gauss-Bonnet term. Because of the
nonlinearity of the field equations, it is very difficult to find
out nontrivial exact analytical solutions of Einstein's equation
with higher curvature terms. In most cases, one has to adopt some
approximation methods or find solutions numerically. These facts
provide a strong motivation for considering new exact solutions of
the Einstein-Gauss-Bonnet gravity with nonlinear source. The main
goal of this work is to present analytical solutions for a typical
class of magnetic horizonless of GB-nonlinear Maxwell source and
investigate their properties. These kinds of work have been
investigated in many papers of Einstein gravity. Static uncharged
cylindrically symmetric solutions of Einstein gravity in four
dimensions were considered in \cite{Levi}. Similar static
solutions in the context of cosmic string theory were found in
\cite{Vil}. All of these solutions \cite{Levi,Vil} are horizonless
and have a conical geometry, which are everywhere flat except at
the location of the line source. An extension to include the
electromagnetic field has also been done \cite{Muk,Lem1}. The
generalization of the four-dimensional solution found in
\cite{Lem1} to the case of $(n+1)$-dimensional solution with all
rotation and boost parameters has been done in \cite{Deh4}.

The outline of our paper is as follows. In next Section, we
briefly present the basic field equations of the GB gravity and
nonlinear Maxwell source. In section \ref{Long}, we present a new
class of static magnetic solutions and consider the properties of
the solutions as well as the energy condition. In section
\ref{Rot}, we endow these spacetime with global rotations and then
apply the counterterm method to compute the conserved quantities
of these solutions. Finally, we finish our paper with some closing
remarks.

\section{Field Equations}
The gravitational and electromagnetic field equations of the
Einstein-GB gravity in the presence of power of Maxwell invariant
field may be written as
\begin{eqnarray}
&&G_{\mu \nu }+\Lambda g_{\mu \nu }-\frac{\alpha }{2}\left[
8R^{\rho \sigma }R_{\mu \rho \nu \sigma }-4R_{\mu }^{\ \rho \sigma
\lambda }R_{\nu \rho \sigma \lambda }-4RR_{\mu \nu }+8R_{\mu
\lambda }R_{\text{ \ }\nu }^{\lambda
}+\right.  \nonumber \\
&&\left. g_{\mu \nu }\left( R_{\mu \nu \gamma \delta }R^{\mu \nu
\gamma \delta }-4R_{\mu \nu }R^{\mu \nu }+R^{2}\right) \right]
=2\kappa \left( sF_{\mu \rho }F_{\nu }^{~\rho
}F^{s-1}-\frac{1}{4}g_{\mu \nu }F^{s}\right) , \label{Geq}
\end{eqnarray}
\begin{equation}
\partial _{\mu }\left( \sqrt{-g}F^{\mu \nu }F^{s-1}\right) =0,  \label{Maxeq}
\end{equation}%
where $G_{\mu \nu }$ is the Einstein tensor, $\Lambda
=-n(n-1)/2l^{2}$ is the negative cosmological constant, $\alpha $
is the GB coefficient with dimension $(\mathrm{length})^{2}$, $R$,
$R_{\mu \nu }$ and $R_{\mu \nu \gamma \delta }$\ are Ricci scalar,
Ricci and Riemann tensors. In addition, $ \kappa $ is a constant
in which we set $\kappa =1$ without loss of generality and
consequently the energy density (the $T_{\widehat{0}\widehat{0 }}$
component of the energy-momentum tensor in the orthonormal frame)
is positive. In the limit $s=1$, the nonlinear electromagnetic
field reduces to the standard Maxwell form, as it should be. It is
easy to show that for $\alpha =0$, the equation (\ref{Geq})
reduces to the Einstein gravity coupled with power Maxwell
invariant source.

\section{Static magnetic branes\label{Long}}

Here we want to obtain the $(n+1)$-dimensional solutions of Eqs.
(\ref{Geq}) and (\ref{Maxeq}) which produce longitudinal magnetic
fields in the Euclidean submanifold spans by $x^{i}$\ coordinates
($i=1,...,n-2$). We will work with the following ansatz for the
metric \cite{Lem1}:
\begin{equation}
ds^{2}=-\frac{\rho ^{2}}{l^{2}}dt^{2}+\frac{d\rho ^{2}}{f(\rho )}%
+l^{2}f(\rho )d\phi ^{2}+\frac{\rho ^{2}}{l^{2}}dX^{2},
\label{Met1a}
\end{equation}
where $dX^{2}={{\sum_{i=1}^{n-2}}}(dx^{i})^{2}$ is the Euclidean
metric on the $(n-2)$-dimensional submanifold. The angular
coordinates $\phi $ is dimensionless as usual and ranges in
$[0,2\pi ]$, while $x^{i}$'s range in $(-\infty ,\infty )$. The
motivation for this metric gauge $[g_{tt}\varpropto -\rho ^{2}$
and $(g_{\rho \rho })^{-1}\varpropto g_{\phi \phi }]$ instead of
the usual Schwarzschild gauge $[(g_{\rho \rho })^{-1}\varpropto
g_{tt}$ and $ g_{\phi \phi }\varpropto \rho ^{2}]$ comes from the
fact that we are looking for a horizonless magnetic solution
instead of electrical one. Also, one can obtain the presented
metric (\ref{Met1a}) with local transformations $ t\rightarrow
il\phi $ and $\phi \rightarrow it/l$ in the horizon flat
Schwarzschild-like metric, $ds^{2}=-f(\rho )dt^{2}+\frac{d\rho
^{2}}{f(\rho ) }+\rho ^{2}d\phi ^{2}+\frac{\rho
^{2}}{l^{2}}dX^{2}$. Thus, the nonzero component of the gauge
potential is  $A_{\phi}$, which can be written as
\begin{equation}
A_{\mu }=-2ql^{n-1}h(\rho )\delta _{\mu }^{\phi },
\end{equation}%
where $h(\rho )$ is $\ln (\rho )$ for $s=n/2$, and for other
values of $s$, we have
\[
h(\rho )=\rho ^{(2s-n)/(2s-1)},
\]
therefore the non-vanishing component of electromagnetic field
tensor is now given by
\begin{equation}
F_{\rho \phi }=2ql^{n-1}\left\{
\begin{array}{cc}
\rho ^{-1}, & s=\frac{n}{2} \\
\frac{2s-n}{2s-1}\rho ^{-(n-1)/(2s-1)}, & \text{Otherwise}%
\end{array}%
\right. .  \label{Ftr}
\end{equation}
Because of vanishing the electromagnetic field for $s=0,1/2$, we
ignore this cases. It is notable that for $s<\frac{1}{2}$, the
electromagnetic field (\ref{Ftr}) diverge as $\rho \longrightarrow
\infty $ and therefore we restrict our solutions to
$s>\frac{1}{2}$. To find the function $f(\rho )$, one may use any
components of Eq. (\ref{Geq}). The solution of Eq. (\ref{Geq}) can
be written as
\begin{equation}
f(\rho )=\frac{2\rho ^{2}}{(n-1)\gamma }\left(
1-\sqrt{1+\frac{2\gamma \Lambda }{n}+\frac{\gamma m}{\rho
^{n}}-\gamma \Gamma (\rho )}\right) , \label{f(r)}
\end{equation}%
\begin{eqnarray}
\Gamma (\rho ) &=&\left\{
\begin{array}{cc}
2^{3n/2}(n-1)l^{n(n-2)}q^{n}\frac{\ln \rho }{\rho ^{n}}, & s=\frac{n}{2} \\
\frac{(2s-1)^{2}}{2s-n}\left( \frac{8l^{2(n-2)}q^{2}(2s-n)^{2}}{%
(2s-1)^{2}\rho ^{2(n-1)/(2s-1)}}\right) ^{s}, & s>\frac{1}{2},s\neq \frac{n}{%
2}%
\end{array}%
\right. ,  \label{GAMMA1} \\
\gamma &=&\frac{4\alpha (n-2)(n-3)}{(n-1)},  \nonumber
\end{eqnarray}%
where mass parameter, $m,$ is related to integration constant. It
is easy to show that for $\alpha \longrightarrow 0$, Eq.
(\ref{f(r)}) reduces to
\begin{equation}
f_{E}(\rho )=\frac{-2\Lambda \rho ^{2}}{n(n-1)}-\frac{m}{(n-1)\rho ^{n-2}}+%
\frac{\rho ^{2}}{(n-1)}\Gamma (\rho ),  \label{Einstein}
\end{equation}%
where $f_{E}(\rho )$ is the Einstein solution of Eq.
(\ref{Geq})($\alpha =0$).

\subsection{Energy conditions}
Here, we discuss the energy conditions for the power Maxwell
invariant electromagnetic field in diagonal metric. For the energy
momentum tensor written in the orthonormal contravariant basis
vectors as $T^{\mu \nu }=diag(
%TCIMACRO{\U{3bc} }%
%BeginExpansion
\mu
%EndExpansion
,p_{r},p_{t_{1}},p_{t_{2}},$\textperiodcentered
\textperiodcentered \textperiodcentered $)$, the null energy
condition (NEC) is the assertion
that $p_{r}+%
%TCIMACRO{\U{3bc} }%
%BeginExpansion
\mu
%EndExpansion
\geq 0$ and $p_{t_{i}}+%
%TCIMACRO{\U{3bc} }%
%BeginExpansion
\mu
%EndExpansion
\geq 0$, and the weak energy condition (WEC) implies $%
%TCIMACRO{\U{3bc} }%
%BeginExpansion
\mu
%EndExpansion
\geq 0$, $p_{r}+%
%TCIMACRO{\U{3bc} }%
%BeginExpansion
\mu
%EndExpansion
\geq 0$, and $p_{t_{i}}+%
%TCIMACRO{\U{3bc} }%
%BeginExpansion
\mu
%EndExpansion
\geq 0$, while the dominant energy condition (DEC) implies $
%TCIMACRO{\U{3bc} }%
%BeginExpansion
\mu
%EndExpansion
\geq 0$, $-%
%TCIMACRO{\U{3bc} }%
%BeginExpansion
\mu
%EndExpansion
\leq p_{r}\leq
%TCIMACRO{\U{3bc} }%
%BeginExpansion
\mu
%EndExpansion
$, and $-%
%TCIMACRO{\U{3bc} }%
%BeginExpansion
\mu
%EndExpansion
\leq p_{t_{i}}\leq
%TCIMACRO{\U{3bc} }%
%BeginExpansion
\mu
%EndExpansion
$, and strong energy condition (SEC)which implies $p_{r}+%
%TCIMACRO{\U{3bc} }%
%BeginExpansion
\mu
%EndExpansion
\geq 0$, $p_{t_{i}}+%
%TCIMACRO{\U{3bc} }%
%BeginExpansion
\mu
%EndExpansion
\geq 0$, and $%
%TCIMACRO{\U{3bc} }%
%BeginExpansion
\mu
%EndExpansion
+p_{r}+\sum_{i=1}^{n-1}p_{t_{i}}\geq 0$. The physical
interpretations of $
%TCIMACRO{\U{3bc} }%
%BeginExpansion
\mu
%EndExpansion
$, $p_{r}$, and $p_{t_{i}}$ are energy density, radial pressure,
and the tangential pressure, respectively. For our diagonal
metric, using the orthonormal contravariant (hatted) basis vectors
\[
\mathbf{e}_{\widehat{t}}=\frac{l}{r}\frac{\partial }{\partial
t},\text{ \ \ } \mathbf{e}_{\widehat{r}}=f^{1/2}\frac{\partial
}{\partial r},\text{ \ \ } \mathbf{e}_{\widehat{\phi
}}=\frac{1}{lf^{1/2}}\frac{\partial }{\partial \phi },\text{ \
}\mathbf{e}_{\widehat{x^{i}}}=\frac{l}{r}\frac{\partial }{
\partial x^{i}},
\]
the mathematics and physical interpretations become simplified. It
is a matter of straight forward calculations to show that the
stress-energy tensor is
\begin{eqnarray}
T_{_{\widehat{t}\widehat{t}}}
&=&-T_{_{\widehat{i}\widehat{i}}}=\frac{1}{2}
\left( \frac{2F_{\phi r}^{2}}{l^{2}}\right) ^{s}, \\
&&  \nonumber \\
\text{ \ \ }T_{_{\widehat{r}\widehat{r}}} &=&T_{_{\widehat{\phi
}\widehat{ \phi }}}=\frac{2s-1}{2}\left( \frac{2F_{\phi
r}^{2}}{l^{2}}\right) ^{s}, \label{EMtensor}
\end{eqnarray}
so for satisfaction of the null and weak energy condition, we
should justify $s>0$.
\begin{equation}
T_{_{\widehat{t}\widehat{t}}}\geq
0,\hspace{1cm}T_{_{\widehat{t}\widehat{t}
}}+T_{_{\widehat{i}\widehat{i}}}\geq 0,\text{ \
}T_{_{\widehat{t}\widehat{t}
}}+T_{_{\widehat{r}\widehat{r}}}=T_{_{\widehat{t}\widehat{t}}}+T_{_{\widehat{
\phi }\widehat{\phi }}}\geq 0.  \label{WEC}
\end{equation}
One may show that for satisfaction of the dominant and strong
energy conditions, we should set $0<s<1$ and $s\geqslant
\frac{n-1}{4}$, respectively. Since for Einstein gravity or (GB
gravity) $n\geqslant 3$ or $(4)$ and also, we restrict our
solutions to $s>\frac{1}{2}$, the presented solutions always
satisfy the null, weak and strong energy conditions,
simultaneously, and dominant energy condition is satisfied when
$\frac{1}{2}<s\leqslant 1$.
\begin{figure}[tbp]
\epsfxsize=10cm \centerline{\epsffile{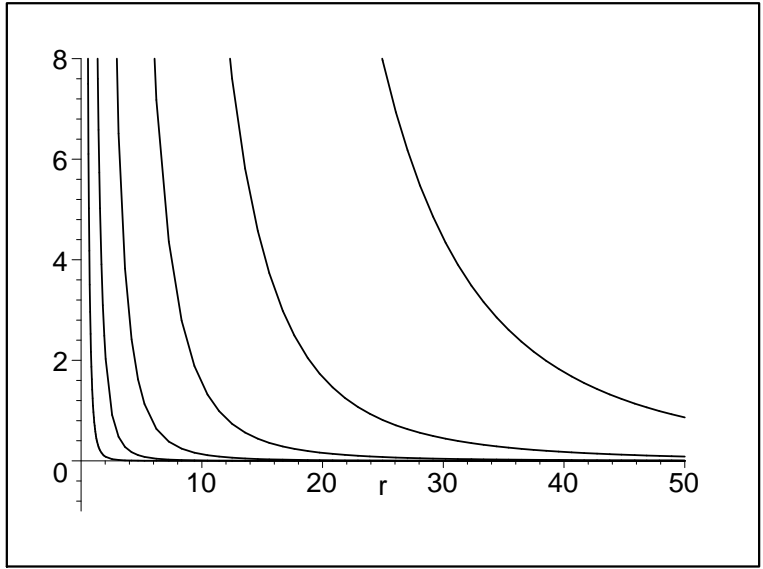}} \caption{The
energy density $T_{_{\widehat{t}\widehat{t}}}$ of power Maxwell
invariant versus $r$ for $n=4$, $l=1$, $q=1$, and $s=3,4,5,6,7,8$
from left to right, respectively.} \label{T00r}
\end{figure}
\begin{figure}[tbp]
\epsfxsize=10cm \centerline{\epsffile{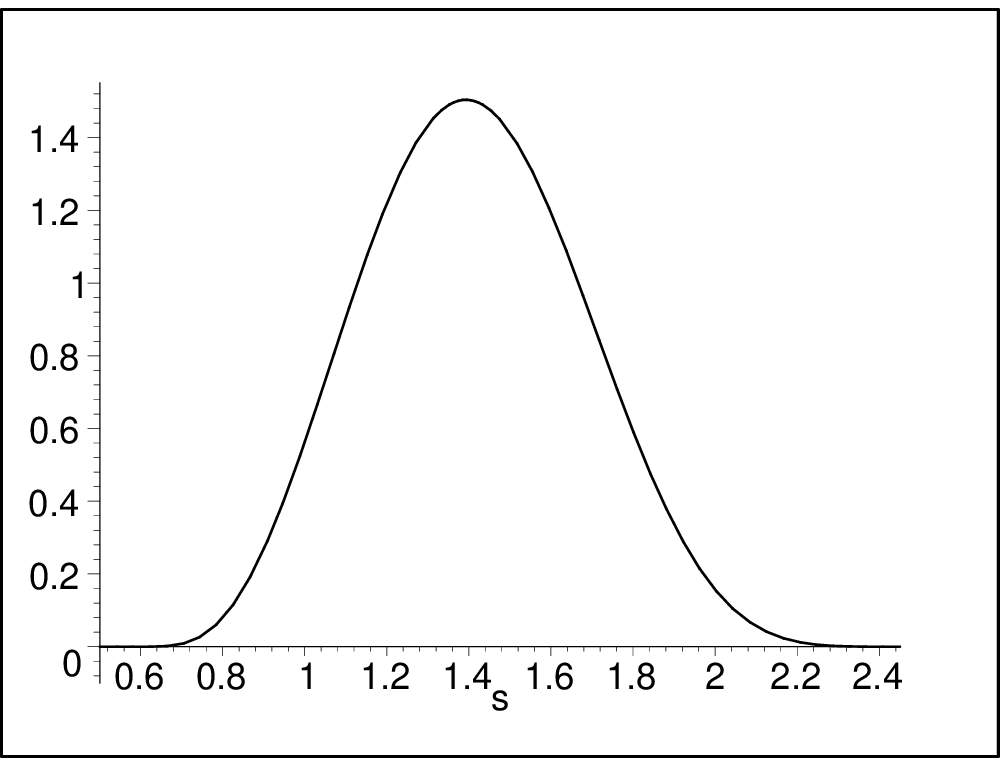}} \caption{The
energy density $T_{_{\widehat{t}\widehat{t}}}$ of power Maxwell
invariant versus $s$ for $n=5$, $l=1$, $q=2$, and $r=2$ for
$\frac{1}{2}<s<\frac{n}{2}$.} \label{T00s1}
\end{figure}
\begin{figure}[tbp]
\epsfxsize=10cm \centerline{\epsffile{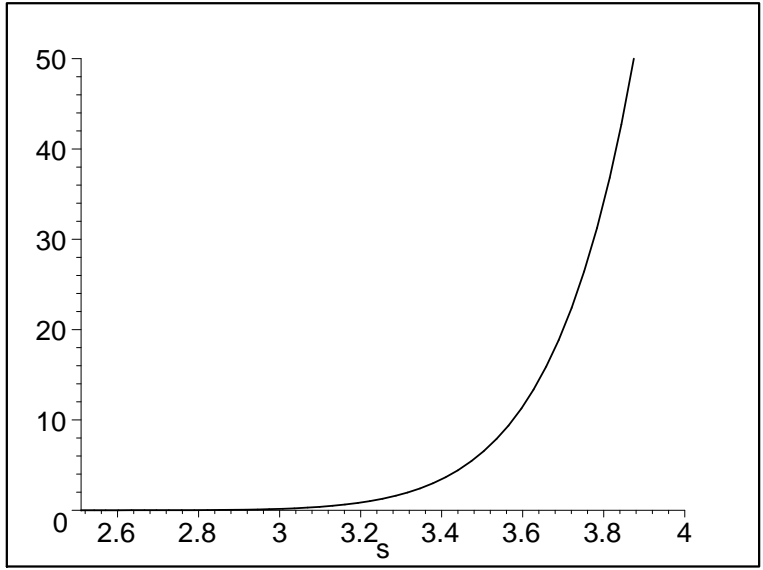}} \caption{The
energy density $T_{_{\widehat{t}\widehat{t}}}$ of power Maxwell
invariant versus $s$ for $n=5$, $l=1$, $q=2$, and $r=2$ for
$s>\frac{n}{2}$.} \label{T00s2}
\end{figure}

In order to investigate the effect of the nonlinearity of the
electromagnetic field on energy density of the spacetime, we plot
the $T_{_{\widehat{t}\widehat{t}}}$ versus $r$ (for different
values of nonlinearity parameter $s$) and $s$. Figs. \ref{T00r}
and \ref{T00s2} show that for $s>\frac{n}{2}$, on one hand, for a
fixed value of $r$, as nonlinearity parameter increases, the
energy density of the spacetime increase too and on the other
hand, in order to reduce the concentration area of the energy
density, we should reduce the nonlinearity parameter. Also, Fig.
\ref{T00s1} shows that $T_{_{\widehat{t}\widehat{t}}}$ has a local
maximum when the nonlinearity parameter changes from $\frac{1}{2}$
to $\frac{n}{2}$.

\subsection{Conformally invariant electromagnetic field\label{Conformally}}
It is easy to show that the clue of the conformal invariance of
Maxwell source lies in the fact that we have considered power of
the Maxwell invariant, $F=F_{\mu \nu }F^{\mu \nu }$. Here we want
to justify the nonlinearity parameter $s$, such that the
electromagnetic field equation be invariant under conformal
transformation ($g_{\mu \nu }\longrightarrow \Omega ^{2}g_{\mu \nu
}$ and $A_{\mu }\longrightarrow A_{\mu }$). The idea is to take
advantage of the conformal symmetry to construct the analogues of
the four dimensional Reissner-Nordstr\"{o}m solutions in higher
dimensions. It is easy to show that for Lagrangian in the form
$L(F)$ in $(n+1)$ -dimensions, $T_{\mu }^{\mu }\propto \left[
F\frac{dL}{dF}-\frac{n+1}{4}L \right] $; so $T_{\mu }^{\mu }=0$
implies $L(F)=Constant\times F^{(n+1)/4}$. For our case, nonlinear
Maxwell field, $L(F)\propto F^{s}$, we should set $s=(n+1)/4$ for
conformally invariance condition. It is worthwhile to mention that
Since $n\geq 3$ and therefore $s=(n+1)/4\geq 1$, one can show that
the magnetic solutions with conformally invariant Maxwell source
are asymptotically AdS in arbitrary dimensions. In this case the
functions $f(\rho )$ and $h(\rho )$ reduce to
\begin{eqnarray}
f(\rho ) &=&\frac{2\rho ^{2}}{(n-1)\gamma }\left(
1-\sqrt{1+\frac{2\gamma \Lambda }{n}+\frac{\gamma m}{\rho
^{n}}+\gamma g(\rho )}\right) ,
\label{f(r)Conf} \\
&&g(\rho )=2^{(n-3)/4}(n-1)\left( \frac{2l^{n-2}q}{\rho
^{2}}\right) ^{(n+1)/2},  \nonumber
\end{eqnarray}
\begin{equation}
h(\rho )\propto \frac{1}{\rho },  \label{Em2}
\end{equation}
and therefore the electromagnetic field is analogues of the four
dimensional Reissner-Nordstr\"{o}m solutions, $F_{\phi \rho
}\propto \rho ^{-2}$ in arbitrary dimensions.

\subsection{The higher dimensional BTZ-like solutions}
The (2+1)-dimensional BTZ solution \cite{BTZ} have obtained a
great importance in recent years because this provide a simplified
model for exploring some conceptual issues, not only about black
hole thermodynamics and magnetic solutions but also about quantum
gravity and string theory \cite{BTZ1}. The line element of BTZ
solution with negative cosmological constant $\Lambda =-1/l^{2}$
may be written as
\begin{equation}
ds^{2}=-f(\rho )dt^{2}+\frac{d\rho ^{2}}{f(\rho )}+\rho ^{2}d\phi
^{2}, \label{BTZmetric}
\end{equation}
where
\[
f(\rho )=-M+\frac{\rho ^{2}}{l^{2}}+\frac{Q^{2}}{2}\ln \rho ,
\]
in which $M$ and $Q$ are the mass and the electric charge of the
solution, respectively \cite{BTZ2}.

The (2+1)-dimensional static subsection of the metric
(\ref{Met1a}) can be written as
\begin{equation}
ds^{2}=-\frac{\rho ^{2}}{l^{2}}dt^{2}+\frac{d\rho ^{2}}{f(\rho )}
+l^{2}f(\rho )d\phi ^{2},  \label{MetBTZ}
\end{equation}
One can obtain the presented magnetic metric (\ref{MetBTZ}) with
local transformations $t\rightarrow il\phi $ and $\phi \rightarrow
it/l$ in the electrical BTZ metric (\ref{BTZmetric}) with the same
metric function $f(\rho )$.

Comparing (\ref{MetBTZ}) with (\ref{Met1a}) help us to conclude
that Eqs. (\ref{Ftr}), (\ref{f(r)}) and (\ref{Einstein}) with
metric (\ref{Met1a}) may be interpreted as higher dimensional
BTZ-like magnetic solutions for $s=\frac{n}{2}$. It is easy to
show that in 3 dimension ($n=2$), the original magnetic BTZ
solution emerge. It is notable that for $s=\frac{n}{2}$, BTZ-like
solutions, the electromagnetic field $F_{\phi \rho }\propto \rho
^{-1}$ in arbitrary dimensions.

\subsection{Properties of the solutions}
At first, we investigate the effects of the nonlinearity on the
asymptotic behavior of the Einstein and GB solutions. It is
worthwhile to mention that for $s>\frac{1}{2}$ (including
$s=\frac{n}{2}$), the asymptotic behavior of
Einstein-(GB)-nonlinear Maxwell field solutions are the same as
Einstein-(GB)-Born-Infeld and linear AdS case.

In order to study the general structure of these spacetime, we
first look for the essential singularity(ies). After some
algebraic manipulation, one can show that for the rotating metric
(\ref{Met1a}), the Kretschmann and Ricci scalars are
\begin{eqnarray}
R_{\mu \nu \rho \sigma }R^{\mu \nu \rho \sigma } &=&f^{\prime
\prime 2}(\rho )+\frac{2(n-1)f^{\prime 2}(\rho )}{\rho
^{2}}+\frac{2(n-1)(n-2)f^{2}(\rho )}{
\rho ^{4}},  \label{RR} \\
R &=&-f^{\prime \prime }(\rho )-\frac{2(n-1)f^{\prime }(\rho
)}{\rho }-\frac{ (n-1)(n-2)f(\rho )}{\rho ^{2}},  \label{R}
\end{eqnarray}
where prime and double prime are first and second derivative with
respect to $\rho $ , respectively. Denoting the largest real root
of $1+\frac{2\gamma \Lambda }{n}+\frac{\gamma m}{\rho ^{n}}-\gamma
\Gamma (\rho )=0$ (in the case that it has real root(s)) by
$r_{1}$, Eq. (\ref{f(r)}) show that $\rho $ should be greater than
$r_{1}$ in order to have a real spacetime. By substituting the
metric function (\ref{f(r)}), It is easy to show that the
Kretschmann invariant and Ricci scalar diverge at
$r_{0}=Max\{0,r_{1}\}$ and they are finite for $\rho >r_{0}$. It
is notable that as $\rho \rightarrow \infty $, we have
\begin{eqnarray}
R_{\mu \nu \rho \sigma }R^{\mu \nu \rho \sigma }
&=&\frac{8(n+1)}{n(n-1)^{2}}
\Lambda ^{2},  \label{RR2} \\
R &=&\frac{2(n+1)}{(n-1)}\Lambda ,  \label{R2}
\end{eqnarray}
which confirm that asymptotic behavior of the solutions is AdS.
Considering the divergency of the Kretschmann and Ricci scalars,
one might think that there is a curvature singularity located at
$\rho =r_{0}$. Two cases happen. In the first case the function
$f(\rho )$ has no real root greater than $r_{0}$, and therefore we
encounter with a naked singularity which we are not interested in
it. So we consider only the second case which the function has one
or more real root(s) larger than $r_{0}$. In this case the
function $f(\rho )$ is negative for $\rho <r_{+}$, and positive
for $\rho >r_{+}$ where $r_{+}$ is the largest real root of
$f(\rho )=0$. This leads to an apparent change of signature of the
metric, and therefore indicates that $\rho $ should be greater
than $r_{+}$. Thus the coordinate $\rho $ assumes the value
$r_{+}\leq \rho <\infty $. The function $f(\rho )$ given in Eq.
(\ref{f(r)}) is positive in the whole spacetime and is zero at
$\rho =r_{+}$, (while $f^{\prime }(\rho =r_{+})\neq 0$). Thus, one
cannot extend the spacetime to $\rho <r_{+}$. To get rid of this
incorrect extension, we introduce the new radial coordinate $r$ as
\[
r^{2}=\rho ^{2}-r_{+}^{2}\Rightarrow d\rho
^{2}=\frac{r^{2}}{r^{2}+r_{+}^{2}} dr^{2}.
\]
With this new coordinate, the metric (\ref{Met1a}) is
\begin{equation}
ds^{2}=-\frac{r^{2}+r_{+}^{2}}{l^{2}}dt^{2}+\frac{r^{2}}{
(r^{2}+r_{+}^{2})f(r)}dr^{2}+l^{2}f(r)d\phi
^{2}+\frac{r^{2}+r_{+}^{2}}{l^{2} }dX^{2},  \label{Metr1b}
\end{equation}
where the coordinate $r$ and $\phi $ assume the value $0\leq r<$
$\infty $ and $0\leq \phi <2\pi $. The function $f(r)$ is now
given as
\begin{equation}
f(r)=\frac{2(r^{2}+r_{+}^{2})}{(n-1)\gamma }\left(
1-\sqrt{1+\frac{2\gamma \Lambda }{n}+\frac{\gamma
m}{(r^{2}+r_{+}^{2})^{n/2}}-\gamma \Gamma (r)} \right) ,
\label{F2}
\end{equation}
where $\Gamma (r)$ changes to
\begin{equation}
\Gamma (r)=\left\{
\begin{array}{cc}
2^{(3n-2)/2}(n-1)l^{n(n-2)}q^{n}\frac{\ln (r^{2}+r_{+}^{2})}{
(r^{2}+r_{+}^{2})^{n/2}}, & s=\frac{n}{2} \\
\frac{(2s-1)^{2}}{2s-n}\left( \frac{8l^{2(n-2)}q^{2}(2s-n)^{2}}{
(2s-1)^{2}(r^{2}+r_{+}^{2})^{(n-1)/(2s-1)}}\right) ^{s}, &
s>\frac{1}{2} ,s\neq \frac{n}{2}
\end{array}
\right. ,
\end{equation}
and $\gamma $ remains unchanged. The electromagnetic field
equation in the new coordinate is
\begin{equation}
F_{r\phi }=2ql^{n-1}\left\{
\begin{array}{cc}
(r^{2}+r_{+}^{2})^{-1/2}, & s=\frac{n}{2} \\
\frac{2s-n}{2s-1}(r^{2}+r_{+}^{2})^{-(n-1)/(4s-2)}, &
s>\frac{1}{2},s\neq \frac{n}{2}
\end{array}
\right. .  \label{f33}
\end{equation}
The function $f(r)$ given in Eq. (\ref{F2}) is positive in the
whole spacetime and is zero at $r=0$. One can easily show that the
Kretschmann scalar does not diverge in the range $0\leq r<\infty
$. However, the spacetime has a conic geometry and has a conical
singularity at $r=0$, since:
\begin{equation}
\lim_{r\rightarrow 0}\frac{1}{r}\sqrt{\frac{g_{\phi \phi
}}{g_{rr}}}\neq 1. \label{limit}
\end{equation}
For more explanations, using a Taylor expansion, in the vicinity
of $r=0$, we can rewrite (\ref{F2})
\begin{equation}
f(r)=f(r)\left\vert _{r=0}\right. +\left( \frac{df}{dr}\left\vert
_{r=0}\right. \right) r+\frac{1}{2}\left(
\frac{d^{2}f}{dr^{2}}\left\vert _{r=0}\right. \right)
r^{2}+O(r^{3})+...,
\end{equation}
where
\begin{eqnarray}
f(r)\left\vert _{r=0}\right. &=&\frac{df}{dr}\left\vert
_{r=0}\right. =0,
\label{ffp} \\
\frac{d^{2}f}{dr^{2}}\left\vert _{r=0}\right. &\neq &0.
\end{eqnarray}

As a result, we can rewrite Eq. (\ref{Metr1b})
\begin{equation}
ds^{2}=-\frac{r_{+}^{2}}{l^{2}}dt^{2}+\frac{2\left(
\frac{d^{2}f}{dr^{2}} \left\vert _{r=0}\right. \right)
^{-1}}{r_{+}^{2}}dr^{2}+\frac{l^{2}}{2} \left(
\frac{d^{2}f}{dr^{2}}\left\vert _{r=0}\right. \right) r^{2}d\phi
^{2}+ \frac{r_{+}^{2}}{l^{2}}dX^{2},
\end{equation}
and since $\frac{d^{2}f}{dr^{2}}\left\vert _{r=0}\right. \neq
\frac{ 2}{lr_{+}}$, one can show that
\begin{equation}
\lim_{r\longrightarrow 0}\frac{1}{r}\sqrt{\frac{g_{\phi \phi
}}{g_{rr}}} =\lim_{r\longrightarrow
0}\frac{r_{+}}{r^{2}}lf(r)=\frac{lr_{+}}{2}\left(
\frac{d^{2}f}{dr^{2}}\left| _{r=0}\right. \right) \neq 1.
\end{equation}
which clearly shows that the spacetime has a conical singularity
at $r=0$ since, when the radius $r$ tends to zero, the limit of
the ratio circumference/radius is not $2\pi $. The canonical
singularity can be removed if one identifies the coordinate $\phi$
with the period
\begin{equation}
\text{Period}_{\phi }=2\pi \left( \lim_{r\rightarrow
0}\frac{1}{r}\sqrt{ \frac{g_{\phi \phi }}{g_{rr}}}\right)
^{-1}=2\pi (1-4\mu ),
\end{equation}
where $\mu$ is given by
\begin{equation}
\mu =\frac{1}{4}\left[ 1-\frac{2}{lr_{+}}\left(
\frac{d^{2}f}{dr^{2}} \left\vert _{r=0}\right. \right)
^{-1}\right] .  \label{mu}
\end{equation}
\begin{figure}[tbp]
\epsfxsize=10cm \centerline{\epsffile{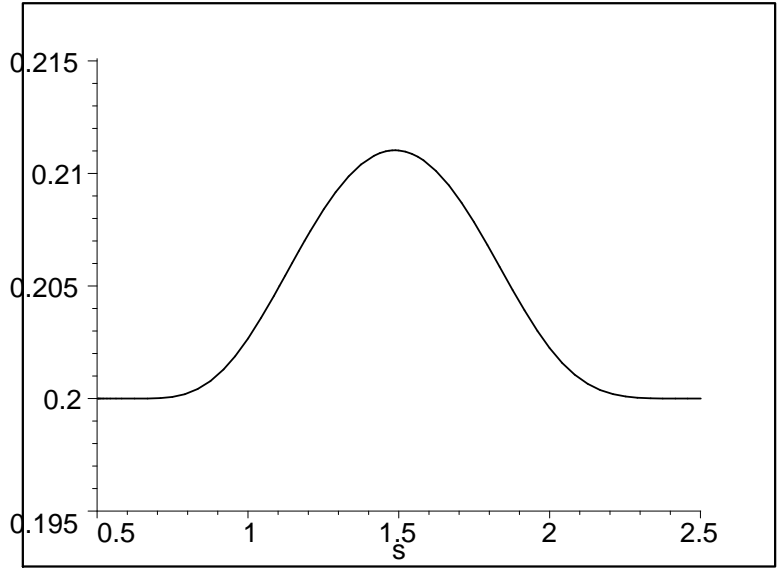}}
\caption{Deficit angle: $\delta \phi/{8\pi}$ versus $s$ for $n=5$,
$l=1$, $q=2$, and $r_{+}=2$ for $\frac{1}{2}<s<\frac{n}{2}$.}
\label{deficit1}
\end{figure}
\begin{figure}[tbp]
\epsfxsize=10cm \centerline{\epsffile{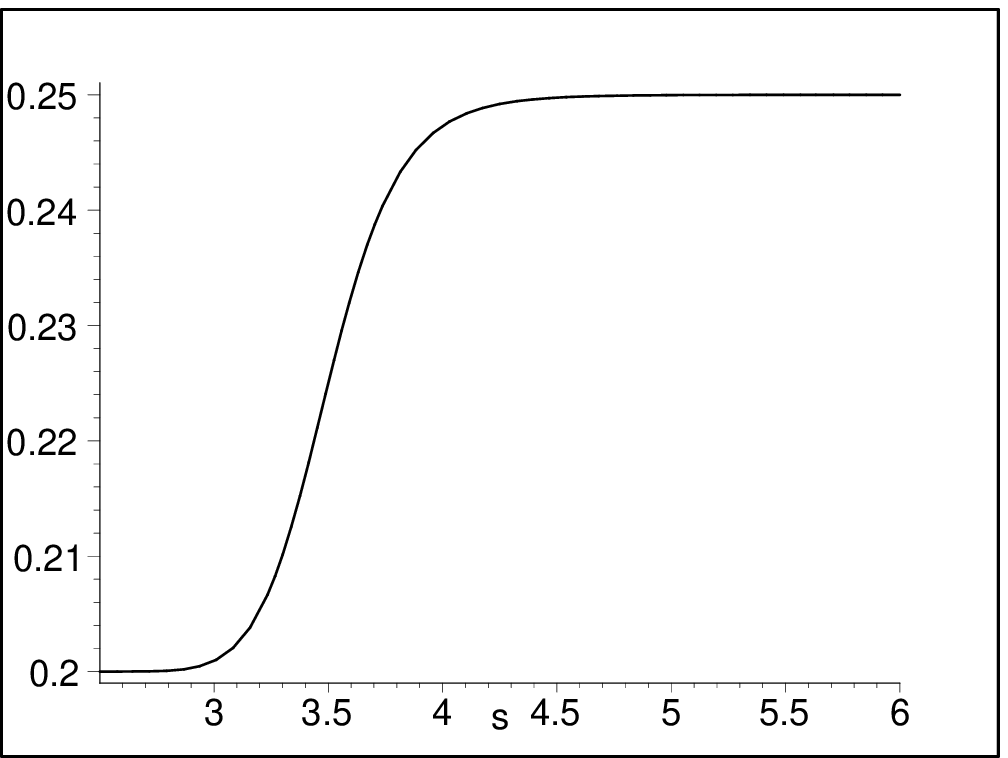}}
\caption{Deficit angle: $\delta \phi/{8\pi}$ versus $s$ for $n=5$,
$l=1$, $q=2$, and $r_{+}=2$ for $s>\frac{n}{2}$.} \label{deficit2}
\end{figure}
\begin{figure}[tbp]
\epsfxsize=10cm \centerline{\epsffile{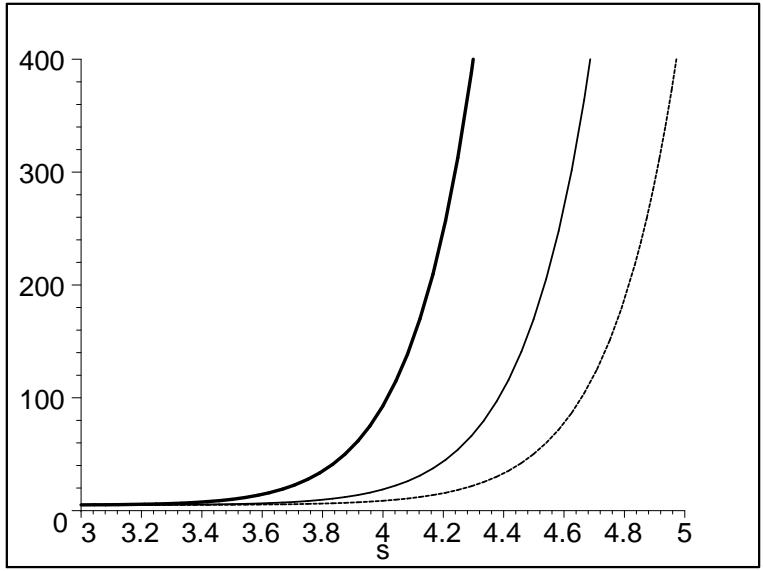}} \caption{$\left(
\frac{d^{2}f}{dr^{2}} \left\vert _{r=0}\right. \right)$ versus $s$
for $n=5$, $l=1$, $q=2$, $\alpha=2$, and $r_{+}=2$ (bold line),
$r_{+}=3$ (continuous line) and $r_{+}=4$ (dashed line) .}
\label{d2F}
\end{figure}
By the above analysis, one concludes that near the origin $r=0$
the metric (\ref{Metr1b}) describes a spacetime which is locally
flat but has a conical singularity at $r=0$ with a deficit angle
$\delta \phi =8\pi \mu $. It is worthwhile to mention that the
magnetic solutions obtained here have distinct properties relative
to the electric solutions obtained in \cite{HendiEslamPanah}.
Indeed, the electric solutions have curvature singularity and
horizon(s) and interpreted as black hole (brane) solutions, while
the magnetic horizonless solutions have conic singularity. In
order to interpreted these solutions, we should mention that near
the origin, this metric in $4$ dimensions is identical to the
spacetime generated by a cosmic string, for which $\mu $ can be
interpreted as the mass per unit length of the string. Thus, here
we may interpret $\mu $ as the mass per unit volume of the brane.
In order to investigate the effect of the nonlinearity of the
magnetic field on $\mu $, we plot the deficit angle $\delta \phi$
versus the nonlinearity parameter $s$. This is shown in Figs.
\ref{deficit1} and \ref{deficit2}, which show that the deficit
angle has a local maximum ($\delta \phi_{m}/{8\pi} \approx
0.2110$) for $\frac{1}{2}<s<\frac{n}{2}$. For $s>\frac{n}{2}$, the
deficit angle is an increasing function, and for large values of
nonlinearity parameter $s$, it goes to an asymptotic value
($\delta \phi_{asy}/{8\pi} \approx 0.2500$). It is easy to show
that for $s=\frac{n}{2}$ with $n=5$, $l=1$, $q=2$, and $r_{+}=2$,
we obtain $\delta \phi_{\frac{n}{2}}/{8\pi}=0.2487$. It is
worthwhile to mention that for arbitrary choice of metric
parameters, we have
\begin{equation}
\delta \phi_{m}<\delta \phi_{\frac{n}{2}}<\delta \phi_{asy}
\end{equation}
\begin{equation}
\lim_{s \longrightarrow \frac{1}{2}^{+}} {\delta \phi}=\lim_{s
\longrightarrow \frac{n}{2}^{-}} {\delta \phi}=\lim_{s
\longrightarrow \frac{n}{2}^{+}} {\delta \phi}\neq {\delta
\phi}|_{s=\frac{n}{2}}
\end{equation}

One can find easily that the function $\left(
\frac{d^{2}f}{dr^{2}} \left\vert _{r=0}\right. \right)$ is an
increasing function of nonlinearity parameter, $s$ (see Fig.
\ref{d2F}). Thus for large values of $s$, this function goes to
infinity, second term in Eq. (\ref{mu}) vanishes, and therefore,
the asymptotic value for $\delta \phi/{8\pi}$ is $0.25$. One may
conclude that since the nonlinearity parameter s, has an effect on
the energy density $T_{_{\widehat{t}\widehat{t}}}$ and the metric
function $f(r)$, so it can directly have an effect on the deficit
angle of conic singularity.

\section{Spinning Magnetic Branes\label{Rot}}
Here, we desire to give rotation to our spacetime solutions
(\ref{Metr1b}). In order to add angular momentum to the spacetime,
we perform the following rotation boost in the $t$-$\phi $ plane
\begin{equation}
t\mapsto \Xi t-a\phi \ \ \ \ \ \ \ \ \ \ \phi \mapsto \Xi \phi
-\frac{a}{l^{2}}t,  \label{tphi}
\end{equation}
where $a$ is the rotation parameter and $\Xi
=\sqrt{1+a^{2}/l^{2}}$. Substituting Eq. (\ref{tphi}) into Eq.
(\ref{Metr1b}) we obtain
\begin{equation}
ds^{2}=-\frac{r^{2}+r_{+}^{2}}{l^{2}}\left( \Xi dt-ad\phi \right)
^{2}+\frac{ r^{2}dr^{2}}{(r^{2}+r_{+}^{2})f(r)}+l^{2}f(r)\left(
\frac{a}{l^{2}}dt-\Xi d\phi \right)
^{2}+\frac{r^{2}+r_{+}^{2}}{l^{2}}dX^{2},  \label{Metr2}
\end{equation}
where $f(r)$ is the same as $f(r)$ given in Eq. (\ref{F2}). The
non vanishing electromagnetic field components become
\begin{equation}
F_{rt}=-\frac{a}{\Xi l^{2}}F_{r\phi }=-\frac{2qal^{n-3}}{\Xi
}\left\{
\begin{array}{cc}
(r^{2}+r_{+}^{2})^{-1/2}, & s=\frac{n}{2} \\
\frac{2s-n}{2s-1}(r^{2}+r_{+}^{2})^{-(n-1)/(4s-2)}, &
s>\frac{1}{2},s\neq \frac{n}{2}
\end{array}
\right. .
\end{equation}
The transformation (\ref{tphi}) generates a new metric, because it
is not a permitted global coordinate transformation. This
transformation can be done locally but not globally. Therefore,
the metrics (\ref{Metr1b}) and (\ref{Metr2}) can be locally mapped
into each other but not globally, and so they are distinct. Again,
this spacetime has no horizon and curvature singularity, However,
it has a conical singularity at $r=0$.

Second, we study the rotating solutions with more rotation
parameters. The rotation group in $n+1$ dimensions is $SO(n)$ and
therefore the number of independent rotation parameters is
$[n/2]$, where $[x]$ is the integer part of $x$. We now generalize
the above solution given in Eq. (\ref{Metr1b}) with $k\leq \lbrack
n/2]$ rotation parameters. This generalized solution can be
written as
\begin{eqnarray}
ds^{2} &=&-\frac{r^{2}+r_{+}^{2}}{l^{2}}\left( \Xi
dt-{{\sum_{i=1}^{k}}} a_{i}d\phi ^{i}\right) ^{2}+f(r)\left(
\sqrt{\Xi ^{2}-1}dt-\frac{\Xi }{\sqrt{
\Xi ^{2}-1}}{{\sum_{i=1}^{k}}}a_{i}d\phi ^{i}\right) ^{2}  \nonumber \\
&&+\frac{r^{2}dr^{2}}{(r^{2}+r_{+}^{2})f(r)}+\frac{r^{2}+r_{+}^{2}}{
l^{2}(\Xi ^{2}-1)}{\sum_{i<j}^{k}}(a_{i}d\phi _{j}-a_{j}d\phi
_{i})^{2}+ \frac{r^{2}+r_{+}^{2}}{l^{2}}dX^{2},  \label{Metr5}
\end{eqnarray}
where $\Xi =\sqrt{1+\sum_{i}^{k}a_{i}^{2}/l^{2}}$, $dX^{2}$ is the
Euclidean metric on the $(n-k-1)$-dimensional submanifold with
volume $V_{n-k-1}$ and $f(r)$ is the same as $f(r)$ given in Eq.
(\ref{F2}). The non-vanishing components of electromagnetic field
tensor are
\begin{equation}
F_{rt}=-\frac{(\Xi ^{2}-1)}{\Xi a_{i}}F_{r\phi
^{i}}=-\frac{2ql^{n-1}(\Xi ^{2}-1)}{\Xi a_{i}}\left\{
\begin{array}{cc}
(r^{2}+r_{+}^{2})^{-1/2}, & s=\frac{n}{2} \\
\frac{2s-n}{2s-1}(r^{2}+r_{+}^{2})^{-(n-1)/(4s-2)}, &
s>\frac{1}{2},s\neq \frac{n}{2}
\end{array}
\right. .
\end{equation}

It is worthful to note that one can find a close relation between
the Kerr-NUT-AdS solutions of Ref. \cite{KerrNUT} and the
presented solutions, Eq. (\ref{Metr5}) with metric function given
in Eq. (\ref{F2}) for vanishing both the Gauss-Bonnet parameter
$\alpha$ and the nonlinearity parameter $s$.

\subsection{Conserved Quantities \label{Conserve}}
Here, we present the calculation of the angular momentum and mass
density of the solutions. Generally, in order to have finite
conserved quantities for asymptotically AdS solutions of Einstein
gravity, one may use of the counterterm method inspired by the
anti-de Sitter/conformal field theory (AdS/CFT) correspondence
\cite{Mal}. In addition, for asymptotically AdS solutions of
Lovelock gravity with flat boundary, $\widehat{R}_{abcd}(\gamma
)=0$ (our solutions), the finite energy momentum tensor is
\cite{DBSH,DM1}
\begin{equation}
T^{ab}=\frac{1}{8\pi }\{(K^{ab}-K\gamma ^{ab})+2\alpha
(3J^{ab}-J\gamma ^{ab})-\left( \frac{n-1}{l_{eff}}\right) \gamma
^{ab}\},  \label{Stress}
\end{equation}
where $l_{eff}$ is
\begin{eqnarray}
l_{eff} &=&3\sqrt{\frac{\zeta }{2}}\frac{\left( 1-\sqrt{1-\zeta
}\right)
^{1/2}}{\left( 1-\sqrt{1-\zeta }+\zeta \right) }l,  \label{L} \\
\zeta &=&\frac{(n-1)\gamma }{l^{2}}.  \nonumber
\end{eqnarray}
It is notable that, when $\alpha $ goes to zero (Einstein
solutions), $l_{eff}$ \ reduces to $l$, as it should be. In Eq.
(\ref{Stress}), $K^{ab}$ is the extrinsic curvature of the
boundary, $K$ is its trace, $\gamma ^{ab}$ is the induced metric
of the boundary, and $J$ is trace of $J^{ab}$
\begin{equation}
J_{ab}=\frac{1}{3}
(K_{cd}K^{cd}K_{ab}+2KK_{ac}K_{b}^{c}-2K_{ac}K^{cd}K_{db}-K^{2}K_{ab}).
\end{equation}
To compute the conserved charges of the spacetime, we should write
the boundary metric in Arnowitt-Deser-Misner form. When there is a
Killing vector field $\mathcal{\xi }$ on the boundary, then the
quasilocal conserved quantities associated with the stress tensors
of Eq. (\ref{Stress}) can be written as
\begin{equation}
\mathcal{Q}(\mathcal{\xi )}=\int_{\mathcal{B}}d^{n-1}\varphi
\sqrt{\sigma } T_{ab}n^{a}\mathcal{\xi }^{b},  \label{charge}
\end{equation}
where $\sigma $ is the determinant of the metric $\sigma _{ij}$,
and $n^{a}$ is the timelike unit normal vector to the boundary
$\mathcal{B}${. }For our case, the magnetic solutions of GB
gravity, the first Killing vector is $\xi =\partial /\partial t$,
therefore its associated conserved charge is the total mass of the
brane per unit volume $V_{n-k-1}$, given by
\begin{equation}
M=\int_{\mathcal{B}}d^{n-1}x\sqrt{\sigma }T_{ab}n^{a}\xi
^{b}=\frac{(2\pi )^{k}}{4}\left[ n(\Xi ^{2}-1)+1\right] m.
\label{Mas}
\end{equation}
For the rotating solutions, the conserved quantities associated
with the rotational Killing symmetries generated by $\zeta
_{i}=\partial /\partial \phi ^{i}$ are the components of angular
momentum per unit volume $V_{n-k-1}$ calculated as
\begin{equation}
J_{i}=\int_{\mathcal{B}}d^{n-1}x\sqrt{\sigma }T_{ab}n^{a}\zeta
_{i}^{b}= \frac{(2\pi )^{k}}{4}n\Xi ma_{i}.  \label{Ang}
\end{equation}
Finally, we calculate the electric charge of the solutions. To
determine the electric field we should consider the projections of
the electromagnetic field tensor on special hypersurfaces. Then
the electric charge per unit volume $V_{n-k-1}$ can be found by
calculating the flux of the electromagnetic field at infinity,
yielding
\begin{equation}
Q=\frac{(2\pi )^{k}}{32}\sqrt{\Xi ^{2}-1}\times \left\{
\begin{array}{cc}
2^{3n/2}l^{n-1}nq^{n-1}, & s=\frac{n}{2} \\
2^{3s+1}l^{2s-1}sq^{2s-1}, & s>\frac{1}{2},s\neq \frac{n}{2}
\end{array}
\right. ,  \label{elecch}
\end{equation}
which show that the electric charge is proportional to the
magnitude of rotation parameters and is zero for the static
solutions ($\Xi =1$).

\section{ Closing Remarks}
In this paper, we started with a new class of static magnetic
solutions in Gauss--Bonnet gravity in the presence of power
Maxwell invariant field. One may obtain this magnetic metric with
transformations $t\rightarrow il\phi $ and $\phi \rightarrow it/l$
in the horizon flat Schwarzschild-like metric. Because of the
periodic nature of $\phi $, this transformation is not a proper
coordinate transformation on the entire manifold. Therefore, the
magnetic and Schwarzschild-like metrics can be locally mapped into
each other but not globally, and so they are distinct \cite{Sta}.
Also, we found that these solutions have no curvature singularity
and no horizon. The metric function $f(r)$ is nonnegative in the
whole spacetime and is zero at $r_{+}$.

Then, we restricted the nonlinearity parameter to $s>1/2$, since
electromagnetic field at spatial infinity should vanish.
Investigation of the energy conditions showed that since $s>1/2$,
the presented magnetic brane solutions satisfied, simultaneously,
the null, weak and strong energy conditions, and only for
$\frac{1}{2}<s \leq 1$, the dominant energy condition satisfied.
Also, we plot the energy density for various $s$, and found that
it has a local maximum when $\frac{1}{2}<s<\frac{n}{2}$, and for
$s>\frac{n}{2}$ it is an increasing function.

In addition, we showed that for a special value of nonlinearity
parameter, $s=(n+1)/4$, the energy--momentum tensor is traceless
and the solutions are conformally invariant. In this case, the
electromagnetic field $F_{\phi r}\propto r^{-2}$ in arbitrary
dimensions and it means that the expression of the Maxwell field
does not depend on the dimensions and its value coincides with the
Reissner-Nordstr\"{o}m solution in four dimension. Also, we
discussed about the special choice of nonlinearity parameter,
$s=n/2$, and interpreted these solutions as higher dimensional
BTZ-like magnetic solutions \cite{BTZlike}. In this case, like BTZ
solutions, the electromagnetic field $F_{\phi r}\propto r^{-1}$ in
arbitrary dimensions.

Then we investigated other properties of the solutions and found
that that for $s>\frac{1}{2}$ (including $s=\frac{n}{2}$), the
asymptotic behavior of Einstein-(GB)-nonlinear Maxwell field
solutions are AdS. Then, we encountered with a conic singularity
at $r=0$ with a deficit angle $\delta \phi$ which is sensitive to
the nonlinearity of the electromagnetic field. We plotted it with
respect to the $s$, and found that, the deficit angle has a local
maximum for $\frac{1}{2}<s<\frac{n}{2}$ and for $s>\frac{n}{2}$,
the deficit angle is an increasing function, and for large values
of nonlinearity parameter $s$, it goes to its asymptotic value,
$\delta \phi=2\pi$.

Calculation of electric charge showed that for the spinning
solutions, when one or more rotation parameters are nonzero, the
solutions has a net electric charge density which is proportional
to the magnitude of the rotation parameter given by $\sqrt{\Xi
^{2}-1}$. This electric charge is sensitive to the nonlinearity
parameter, as it should be.

Finally, we calculated the conserved quantities of the magnetic
branes such as mass, angular momentum and found that these
conserved quantities do not depend on the nonlinearity parameter
$s$. This can be understand easily, since at the boundary at
infinity the effects of the nonlinearity of the electromagnetic
fields vanish (since $s>\frac{1}{2}$).

\section*{Acknowledgements}
This work has been supported financially by Research Institute for
Astronomy and Astrophysics of Maragha, Iran.

\end{document}